\documentclass[doublecol]{epl2}  % for review and submission

\usepackage[utf8]{inputenc}

\usepackage[english]{babel}
\usepackage{graphicx}  
\usepackage{amssymb}   
\usepackage{amsmath}

\DeclareMathOperator{\Tr}{Tr}
\newcommand{\ket}[1]{\left|#1\right\rangle}

\newcommand{\ketbra}[2]{\left|#1\middle\rangle\middle\langle#2\right|}

\newcommand{\be}{\begin{equation}}
\newcommand{\ee}{\end{equation}}
\newcommand{\benn}{\begin{equation*}}
\newcommand{\eenn}{\end{equation*}}

\newcommand{\beq}{\begin{eqnarray}}
\newcommand{\eeq}{\end{eqnarray}}

\def\H1{\widehat{H}_1}

\title{Liouville coherent states}
\author{Matouš Ringel\thanks{E-mail: \email{matous.ringel@unifr.ch}} \and  Vladimir
Gritsev \thanks{E-mail: \email{vladimir.gritsev@unifr.ch}}}
\institute{Department of Physics, University of
Fribourg, Chemin du Musée 3, 1700 Fribourg, Switzerland}


\abstract{For a certain class of open quantum systems there exists a dynamical symmetry
which connects different time-evolved density matrices.  We show how to use this
symmetry for dynamics in the Liouville space with time-dependent parameters.
This allows us to introduce a concept of generalized coherent states 
in the Liouville space (i.e. for density matrices). Dynamics of this class of density
matrices is characterized by  robustness with respect to any time-dependent
perturbations of the couplings.  We study their dynamical context while focusing
on common physical situations corresponding to compact and non-compact
symmetries.}

\begin{document}
\maketitle

\section{Introduction} 
The concept of coherent states plays a very
important role in quantum physics. Introduced by
Schr\"{o}dinger for quantum harmonic oscillator \cite{Sch}, it
received  further application with the birth
of quantum optics in the 60's \cite{QO1,QO2}. Further developments of the
concept of coherent states are associated with the 
generalized coherent states defined for any Lie algebra, with the most
important examples given by the $su(2)$ and $su(1,1)$ algebras.
Coherent states have extremely wide applications in physics and
mathematics, reviewed e.g. in \cite{Klauder,Perelomov,ZFG}.

There are several definitions of the coherent states. 
The oscillator coherent state can be defined either as (i) a state
which minimizes an uncertainty relation, or (ii) an eigenstate of
an annihilation operator, or (iii) a state obtained from the vacuum
state by the action of the displacement operator. These three
definitions are equivalent for the harmonic oscillators while they
are not for the more general (generalized) coherent states
associated to some non-abelian algebras, other than a Heisenberg algebra. The
construction of coherent states associated to
a Lie algebra includes three ingredients: (i) an algebra $g$ with
the representation space; (ii) a vacuum state of ladder
operators in this representation space. This state is supposed to
have an invariant subgroup defined by some subalgebra $h\in g$;
(iii) a concrete representation of a group $G$. It follows from this
that generalized coherent states have profound geometrical
interpretation: they are labeled by the points of a homogeneous
(coset) space $G/H$ which in many important physical situations has
a structure of a K\"{a}hler manifold.

Those quantum-mechanical systems which have a {\it dynamical}
symmetry given by $G$ (e.g. the generators of $G$ commute with the
Schr\"{o}dinger operator $i\partial_{t}-H$, and not just with the
Hamiltonian) can be described by a {\it classical} dynamical system
on a coset $G/H$. 
One of the most important
properties of the generalized coherent states exists due to their Lie
group structure: if the initial state of a quantum system is a
(generalized) coherent state, it will remain so at any subsequent
evolution time (see eg.~\cite{GalitskiLie}). This key property will be important for the
construction below.

The developments briefly outlined above concern the situation of
isolated quantum systems. In this letter we propose a construction scheme of
generalized coherent states for a certain class of quantum systems
with {\it dissipation}. The most convenient 
formulation is in terms of the Liouville space on which the dynamics of
the reduced density matrix is defined by the Liouville operator ${\cal L}$. 
The dynamical symmetry in this case is associated with a symmetry of the density
matrix, so that there is an algebra $G$ of operators commuting with
$\partial_{t}-{\cal L}$. We thus define generalized coherent states
for an open system as  {\it Liouville} coherent states (LCS). 
These states allow for a new interpretation of the Liouville dynamics and
simplify the calculation of typical quantities of interest.
In this paper we first introduce the very concept of LCS and 
then we demonstrate it on several important examples: the simplest
non-trivial ones when the algebra $G$ is isomorphic to $su(2)$ algebra, which
happens for spin-boson-type models, or $su(1,1)$ algebra, which corresponds to
models of harmonic oscillator type. A defining property of these density matrices
is a robustness of evolution with respect to any time-pendent driving.

\section{Dynamical symmetry approach to open quantum systems}
In general, the formal solution of the evolution equation for an
open system is given by the time-ordered exponent
$\rho(t)={\cal T}\exp({\cal L}_{I}(t))\rho(0)$,
where ${\cal L}_{I}$ is a Liouville superoperator (that is, it acts on operators)
evaluated in the interaction picture. When focusing on a subsystem, the reduced
density matrix is evolved in time by the Lindblad generators which are constructed
using the eigenoperators of the subsystem according to the usual technique (see
e.g.~\cite{Puri,CriticalStudyOfMarkov}). It is possible that these superoperators fulfill some
Lie-(super)-algebra identities \cite{book}. This is the starting point of our
construction. We demonstrate that a large class of physically relevant systems
indeed satisfies this assumption.

We consider an evolution of the reduced density matrix in the Lindblad form
\beq\label{LindEq} 
{\cal L}(t) \rho(t)
&=&-\frac{i}{\hbar}[H(t),\rho(t)]\\
&+&\sum\gamma_{j}(t)(2A_{j}\rho(t)A^{\dag}_{j}-A^{\dag}_{j}A_{j}\rho(t)-\rho(t)A^{\dag}_{j}A_{j})\nonumber,
\eeq which is the only form compatible with the positivity of the density
matrix. Here $A_{j}$ are eigenoperators of the subsystem.  In general,
$\gamma(t)=\int_{0}^{t}d\tau Re[Y(t-\tau)]$, where
$Y(x)=\int_{0}^{\infty}d\omega \omega^{-1}\rho(\omega)\exp(-i\omega x)$ is
related to the spectral density $\rho(\omega)$ which contains information about
sub-system-reservoir coupling.  Introducing the following notation: $(A\otimes
B)\rho:= A\rho B$, one defines the action of the individual terms in
(\ref{LindEq}) on the density matrix as $L_{i}\rho$ (e.g. $L_{1}\rho=A\rho
A^{\dag}$, $L_{2}\rho=A^{\dag}A\rho$, $L_{3}\rho=\rho AA^{\dag}$, etc.).

Now we {\it assume} that the components of the Lindblad operators 
obey Lie-algebra type relations
\beq\label{Lalg}
~[L_{i},L_{j}]\rho=f_{ijk}L_{k}\rho,
\eeq
where $f_{ijk}$ are some numbers (structure coefficients). Our
approach is also valid if $f_{ijk}$ are proportional to some operator
which commutes with all the other operators. 
The assumption~\eqref{Lalg} is satisfied for a surprisingly large number 
of systems, 
basically for all the standard Liouvillians discussed in literature,
and it was shown to be a rather general property of 
Lindblad-type dynamics~\cite{book}.

Provided that the relations \eqref{Lalg} are satisfied, the time-ordered
exponent for {\it any} time-dependent coefficients $\gamma_{j}(t)$ is an element
of a Lie group corresponding to the Lie algebra of superoperators. Using the
disentangling technique (see e.g.~\cite{Puri}), the time-ordered exponential can
be transformed into a product of ordinary exponentials
\begin{equation}
    \rho(t)=\prod_j\exp(f_{j}(t) L_{j})\rho(0).
    \label{eq:RhoEvolSuperoperator}
\end{equation}
The relation between the functions~$f_{j}(t)$ and the functions
$\gamma_{j}(t)$ can be derived easily. 
In particular, for the simplest case of the $su(2)$ and $su(1,1)$
algebraic structure, defined by the commutation  relations
$[L_{-},L_{+}]=2\sigma L_{0}$, $[L_{0}, L_{\pm}]=\pm L_{\pm}$, where
$\sigma=\pm1$ refer to the $su(1,1)$ and $su(2)$ cases, 
the function $f_{+}$ can be shown to satisfy the following Riccati-type equation
($j\in \{+,z,-\}$),
\beq\label{eq:Ric}
\dot{f}_{+}-\gamma_{z}(t)f_{+}-\sigma\gamma_{-}(t)f^{2}_{+}-\gamma_{+}(t)=0,
\eeq
while 
the remaining
functions are determined as
$f_{z}(t)=\int_{0}^{t}(\gamma_{z}(t)+2\sigma\gamma_{-}(t)f_{+}(t))dt$
and $f_{-}(t)=\int_{0}^{t}\gamma_{-}(t)\exp(f_{z}(t))dt$. 
This equation can be solved (either analytically or numerically) 
for any functional form of $\gamma_{\pm}(t),\gamma_{z}(t)$, in particular for the case when they do
not depend on time. 
 The solution in such a case reads
\beq
f_{\pm}(t)&=&\frac{(\gamma_{\pm}/D)\sinh(tD)}{\cosh(tD)-(\gamma_{0}/2D)\sinh(tD)},\nonumber\\
f_{0}(t)&=&[\cosh(tD)-(\gamma_{0}/2D)\sinh(tD)]^{-2},
\label{eq:SolutionToRicatiTimeIndependent}
\eeq
where $D=((\gamma_{0}/2)^{2}-\sigma\gamma_{+}\gamma_{-})^{1/2}$. 

The disentangled form~\eqref{eq:RhoEvolSuperoperator} allows  
a direct evaluation of any dynamical quantity of interest, such as
the entropy $S=-\Tr(\rho\log\rho)$ or the purity of the density
matrix $\Tr(\rho^{2})$,  for arbitrary time-dependent couplings.

The use of the Lie group structure greatly simplifies computations of
any correlation functions. For example, the quantity~$Tr(A\rho(t))$ 
with $A\in g$ can be reformulated as
$\partial_{\lambda}Tr(\exp(\lambda A)\rho(t))|_{\lambda=0}$;  the expression
in the brackets can be evaluated using the group multiplication
rules \cite{Perelomov}.

\section{Liouville coherent states (LCS)}
We proceed with the
Liouville space formulation. A Liouville space ${\mathcal L}$ is
defined as a direct product of two Hilbert spaces, ${\mathcal
L}={\cal H}\otimes\tilde{\cal H}$, corresponding to the left and
right vectors in terms of superoperator notations. Thus, a vector
$\ket{A}\in{\cal L}$ corresponds to an operator as follows $\ket{A}=\sum
A_{mn}\ket{m,n} \Leftrightarrow \sum_{m,n} A_{mn} \ketbra{m}{n}$,
where the sum runs over orthonormal bases $\ket{m}$ and $\ket{n}$
of the spaces ${\cal H}$ and $\tilde{\cal H}$ of dimensions  $d,\tilde{d}$. 
To define a geometry on the Lioville space we use the usual scalar product
defined by~$(\hat A, \hat B ) :=
\Tr[\hat A^\dagger B]$ (cf.~\cite{Puri}). This scalar product allows us to introduce the
bra-vectors for each ket-vector in the Liouville space.
The equation
for the density matrix $\dot{\rho}=-i[H,\rho]$ gets converted in
this notation into $\ket{\dot{\rho}}=-i\hat{H}\ket{\rho}$
where $\hat{H}=H-\tilde{H}$. This map is known as a
Choi-Jamio{\l}kowski map between states and
operators and it may be used to extend certain concepts known for
states in the Hilbert space to  states in the Liouville space, 
which correspond to operators in the Hilbert space: the operators
$A=\sum_{j,k}a_{j,k}\ketbra{j}{k}$ are mapped to states
$|A\rangle=\sum_{j,k}a_{j,k}\ket{j,k}$. 

{\bf   Definition of LCS: } We define a Liouville coherent state as a generalized coherent
state in the Liouville space. The operators creating generalized
coherent states thus act between two Hilbert spaces constituting the
Liouville space. The general condition that the initial density matrix
be in the class of coherent states means that some linear combination
of generators of the Lie algebra annihilates 
the state~$\sum_{i}\xi_{i}L_{i}|\rho\rangle =0$
and thus defines a stationary subgroup $H$. We explicitly demonstrate this 
for the $SU(2)$ and $SU(1,1)$ cases, in which $H=U(1)$. 

\section{Geometry of dissipative dynamics}
Since generalized coherent
states are labeled by the points of the cosets $G/H$, where $H$ is a
stationary subgroup, the geometry of this coset space has profound
influence on the dynamics. The main property of this type of
dynamical evolution can be formulated as follows: if the Liouville
operator is an element of the Lie algebra of superoperators, and if
the initial density matrix is a coherent state (in the sense that the
corresponding vector in the Liouville space is a generalized
coherent state), irrespectively of the precise form of the
time-dependent coefficients $\gamma_j(t)$  the subsequent dynamics
will remain in this subclass of the states (density matrices). This
is a direct consequence of the group multiplication property. The
overlap between two such LCS is given by the eigenfunctions of the Laplace-Beltrami operator on
$G/H$ \cite{Perelomov}.  The distance function on this space is a measure of Lodschmidt echo
\cite{Peres,Gorin}. The
area 2-form on $G/H$ corresponds to the imaginary part of the metric and describes the geometric
(Berry)
curvature of the quantum evolution. Generally, the topology of the
coset $G/H$ can be nontrivial.

\section{Examples}
We consider two types of examples which have  great
physical importance. One example refers to the case when the
Liouville operator is a combination of generators of a compact
group, $SU(2)$, while the other example is a realization of a
non-compact, $SU(1,1)$ group. For both $su(2)$ and $su(1,1)$ algebras the coherent state is built upon 
a basis state (which we denote by $\ket{0}$) using a shift-operator~$D(\zeta)$,
such that $\ket{\zeta} :=D(\zeta) \ket{0}$, and 
\begin{equation}
\ket{\zeta} :=e^{\zeta L_+} e^{\eta L_0}e^{-\zeta^* L_-}
\ket{0}  = e^{\xi L_+ - \xi^* L_-} \ket{0},
\label{eq:DefinitionOfCohState}
\end{equation}
where $\eta = -\sigma \log\left( 1 - \sigma |\zeta|^2 \right)$ and the relation 
between $\xi$~and~$\zeta$ reads $\zeta = \tanh|\xi| e^{i\varphi}$ (with $\xi=|\xi|
e^{i\varphi}$) for $su(1,1)$ and
$\zeta = - \tan(\Theta/2) e^{-i\varphi}$ (with $\xi = -(\Theta/2) e^{-i\varphi}$) for
$su(2)$.
Modifying  the arguments of~\cite{Perelomov} for the present case, we may write down the resolution of
identity on the Liouville space in the following form:
\begin{equation}
    \hat 1 = \sum_k \int  d\mu_k(\zeta) \ketbra{k,\zeta}{k,\zeta},
    \label{eq:ResolutionOfUnity}
\end{equation}
where $k$~runs over all representations of the underlying dynamical symmetry group and
$\mu_k(\zeta)$~is the group measure~\cite{Perelomov}.

The action of the evolution superoperator on the coherent state can be easily calculated
starting from its disentangled form~\eqref{eq:RhoEvolSuperoperator}. 
We introduce the operator 
\begin{equation}
    \mathcal{O}  := e^{f_{+}L_{+}} e^{f_{0}L_{0}} e^{f_{-}L_{-}} e^{\xi L_+ - \xi^*
    L_-},
\end{equation}
which can be rewritten in a product form 
%$\mathcal{O} = e^{g_{+}L_{+}} e^{g_{0}L_{0}} e^{g_{-}L_{-}}
%= D(g_+) \cdot e^{( g_{0} - \eta
%    ) L_{0}} \cdot 
%    \exp( \left[g_{-} + g_+^* e^{g_0-\eta}\right]L_{-})$,
\begin{multline}
\mathcal{O} = e^{g_{+}L_{+}} e^{g_{0}L_{0}} e^{g_{-}L_{-}}
\\= D(g_+) \cdot e^{( g_{0} - \eta
    ) L_{0}} \cdot 
    \exp( \left[g_{-} + g_+^* e^{g_0-\eta}\right]L_{-}),
\end{multline}
in which
the functions~$g_i$ are governed by equations
\begin{gather}
g_+ = f_+ + \zeta e^{f_0}\Lambda^{-1}, \quad 
e^{g_0} = (1-\sigma|\zeta|^2)^{-\sigma} e^{f_0}\Lambda^{-2},\nonumber\\
g_- =  -\zeta^* + (1-\sigma|\zeta|^2)^{-\sigma}f_-\Lambda^{-1}, 
\end{gather}
where $\Lambda=1 - \sigma f_- \zeta$. 
As expected, the evolution of an initial coherent
state~$\ket{\zeta}$ leads to a coherent state~$\ket{g_+}$, described by an equation
\begin{equation}
    \mathcal{T}\exp{\mathcal{L}(t)} \ket{\zeta} = c(t, g_0,g_+) \ket{g_+},
    \label{eq:CoherentStateEvolution}
\end{equation}
with the prefactor~$c(t, g_0,g_+) = \bar{c}(t) \cdot e^{(g_0-\eta) k}$,
where $k$~is a representation-dependent constant, defined by~$L_0\ket{0} = k \ket{0}$, and the
factor~$\bar{c}(t)$ is determined solely by the dynamics
within the~U(1) sector. Note that the relation between~$g_+$ and~$\zeta$ is a Möbius
transform, under which circles in the complex plane of~$\zeta$ are mapped onto circles in the
plane of~$g_+$:
a circle~$\zeta = |\zeta| e^{i\varphi}$ is mapped to a circle
of radius~$R$ and origin~$z$ given by 
\begin{gather}
    R = \frac{|\zeta| |e^{f_0}|}{1-|f_-|^2 |\zeta|^2},\quad
    z = f_+ + \sigma e^{f_0} \frac{|\zeta|^2 f_-^*}{1-|f_-|^2|\zeta|^2}.
    \label{eq:CoherentCircle}
\end{gather}
Consideration of a mapping of circles is a convenient way to think about the 
evolution of coherent state parameters. Namely, one can see that while the
radius of a circle remains approximately proportional to the radius of the
initial circle, the origin shifts to some finite value; 
therefore the ``average'' value of a coherent state parameter at a later time is
shifted from zero.

\section{Spin-boson-type models: $SU(2)$ LCS} 
For
the case of spin-boson-like models in the RWA the system is described by a Lindblad equation
of the form~\eqref{LindEq}, in which we set~$H(t) = \sigma^z \cdot \Omega(t)/2$, 
$A_+ = \sigma_+$, $A_- = \sigma_-$, $\gamma_+(t) = \gamma(t) {\bar n}$ and $\gamma_-(t) = \gamma(t)
({\bar n}+1)$,
with $\bar{n}=(\exp(\hbar\Omega/k_{B}T)-1)^{-1}$ and $\gamma(t)$ being system-specific.
Defining 
%$L^{+}=\sigma^{+}\rho\sigma^{-}$, $L^{-}=\sigma^{-}\rho\sigma^{+}$, $L^{0}=\frac{1}{4} \left(\sigma^{z}\rho+\rho\sigma^{z}\right)$,
%$R =\frac{1}{2}\left( \sigma^{z}\rho-\rho\sigma^{z} \right)$,
\begin{gather}
    L^{+} \rho=\sigma^{+}\rho\sigma^{-}, \quad L^{-} \rho=\sigma^{-}\rho\sigma^{+},\nonumber\\
    L^{0}\rho=\frac{1}{4} \left(\sigma^{z}\rho+\rho\sigma^{z}\right), \quad
    R\rho =\frac{1}{2}\left( \sigma^{z}\rho-\rho\sigma^{z} \right),
\end{gather}
and using the fact that
$\sigma^{\pm}\sigma^{\mp}=\frac{1}{2}(1\pm\sigma^{z})$, we obtain the
following algebra:
\begin{gather}
    [L^+,L^-]=2L^0,\quad [L^0,L^\pm]=\pm L^\pm\nonumber\\
    [L^{\pm},R]=[L^{0},R]=0,
\end{gather}    
which 
has a structure of the direct product $su(2)\times
u(1)$ and consequently the supergroup operator can be disentangled
using the procedure outlined above. Therefore the formal operator
solution of the evolution equation
$\dot{\rho}(t)={\cal L}\rho(t)$, with 
%${\cal L}=-i\Omega(t) R+\gamma(t)\bar{n}L^{+}+\gamma(t)(\bar{n}+1)L^{-}
%-\gamma(t)L^{0}-\frac{1}{2}\gamma(t)(2\bar{n}+1)$,
\begin{multline}
{\cal L}=-i\Omega(t) R+
\gamma(t)\bar{n}L^{+}+\gamma(t)(\bar{n}+1)L^{-}\\
-\gamma(t)L^{0}-\frac{1}{2}\gamma(t)(2\bar{n}+1),
\end{multline}
reads 
\begin{equation}
    \rho(t)=e^{f_{+}L^{+}} e^{f_{0}L^{0}} e^{f_{-}L^{-}} e^{-i R \int
\Omega dt } e^{- ({\bar n}+1/2) \int \gamma dt}\rho(0),
\end{equation}
where $f_{\pm,z}(t)$ satisfy the Riccati
equation (\ref{eq:Ric}).

Any matrix, namely the density matrix can be decomposed into irreducible
parts classified by the Casimir operator
$\vec{L}^2 = \frac{1}{2}\left( L^+ L^- + L^- L^+ \right) + L^z L^z$, which can be explicitly
expressed as
$\vec{L}^2 = \frac{3}{8} \left( \rho + \sigma_z \rho \sigma_z \right)$. In fact, using the basis of
$2\times 2$ matrices $\sigma_0\equiv\vec{1}$, $\sigma_\pm$, $\sigma_z$, we can 
write~$\rho = \rho_{j=0} + \rho_{j=1/2}$, where $\rho_{j=0} = \rho_+ \sigma_+ + \rho_- \sigma_-$ and
$\rho_{j=1/2} = \rho_{\uparrow} (1+\sigma_z) + \rho_{\downarrow} (1-\sigma_z)$.
The most general density matrix~$\rho$ for a $su(2)$ system with spin~$1/2$ can be written as 
a linear combination of spin coherent states $ \rho = c_0 \ket{0,+} + c_0^*
\ket{0,-} + c_{1}(\zeta) \ket{1/2; \zeta}, \quad \zeta \ge 0$,
where~$\ket{0,\pm}$ corresponds to~$\sigma_\pm$ (the subspace with $j=0$) and
$\ket{1/2;\zeta}$~is a coherent state with parameter~$\zeta$ built upon the
basis state~$\ket{j=1/2,m=-1/2}$, corresponding to~$(1-\sigma_z)/2$. 
The evolution operator acts independently on the two subspaces.
Obviously, the coherent-state structure on the subspace~$j=0$ is irrelevant, the
only nontrivial evolution there being the evolution of the coefficients~$c$.
From the condition $\Tr\rho = 1$ we derive
 $c_1(\zeta)  =  \sqrt{1+|\zeta|^2}/(1+\zeta)$.

For the sake of a particular example, we  follow~\cite{Anastopoulos} and consider a two-level system with level
splitting~$\omega$, interacting with a one-mode bath of
frequency~$k=\omega+\Delta$. The non-Markovian evolution can be treated by a Lindblad
equation, however with time-dependent terms. A time-dependent energy splitting~$\Omega(t) = \Im \mathcal X(t)$ and a
time-dependent dissipation~$\gamma(t) =  \Re \mathcal X(t)$ are given by 
\beq
\mathcal X (t) = i \left( \omega - \frac{g'}2 \right)
    \frac{ r s  + e^{i\Delta' t} }
    { 
    s + e^{i\Delta' t} 
    },
    \label{eq:SU2Xt}
\eeq
where $\Delta'=\sqrt{\Delta^2+g^2}$, $g'=\Delta'-\Delta$, $s=g'/(\Delta+\Delta')$ and $r=[\omega +
(\Delta+\Delta')/2 ]/ [\omega + (\Delta-\Delta')/2]$; we also set~$\bar{n}=0$. 
As $\gamma_+$~is absent, we have~$f_+=0$, and the solution to the corresponding Riccati equations 
simplifies greatly: $f_-(t) = -f_0(t) = \int^t_0 \gamma(t) dt := \Gamma(t)$.
The time evolution of a  coherent state can be deduced
from~Eq.~\eqref{eq:CoherentStateEvolution}.

\begin{figure}[h]
    \begin{center}
        \includegraphics[width=0.47\textwidth]{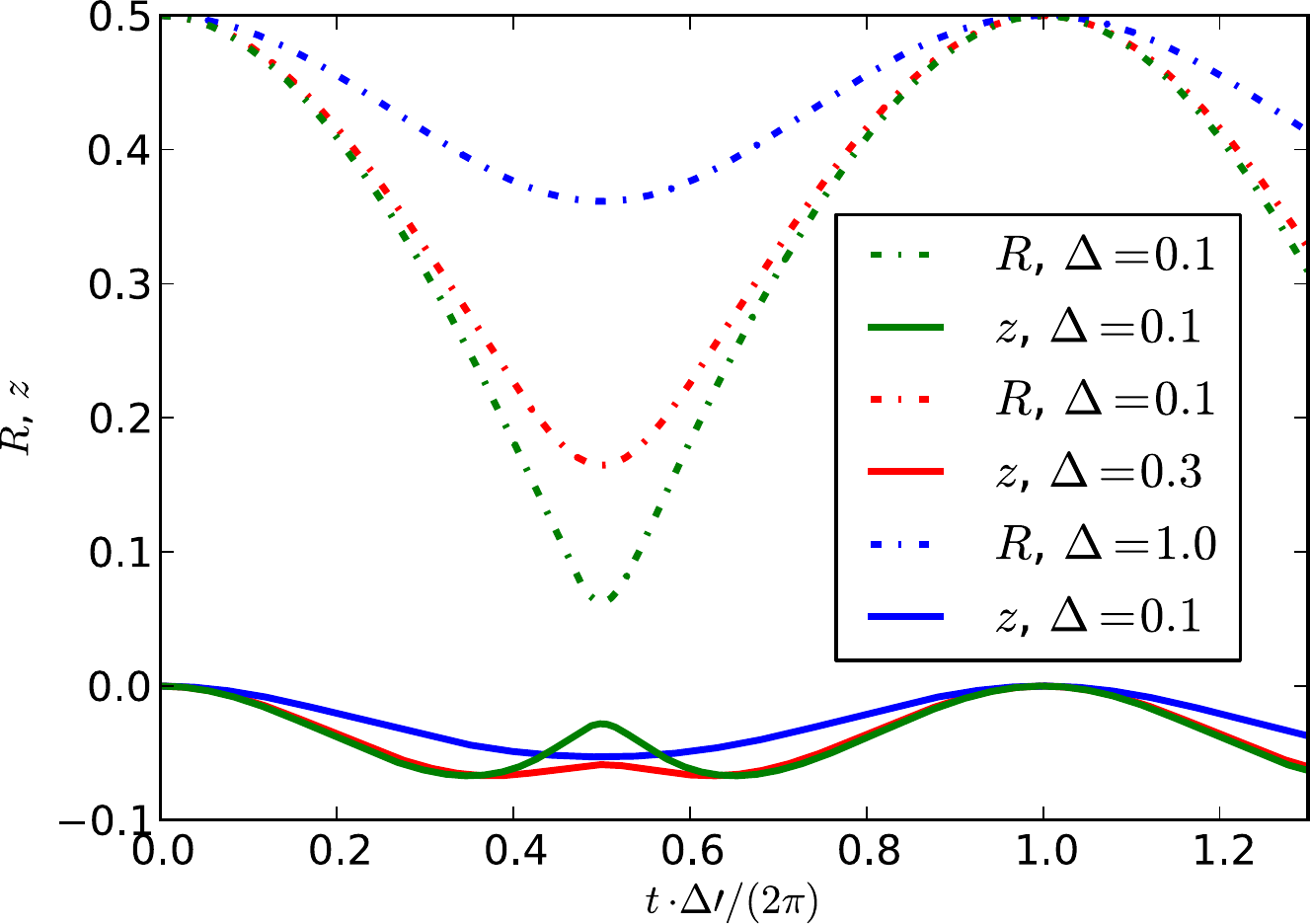}
    \end{center}
    \caption{
    Time dependence of the parameters $R$ and $z$, introduced in
    Eq.~\eqref{eq:CoherentCircle}, for the $su(2)$-type model defined by
    Eq.~\eqref{eq:SU2Xt} (with parameters $\omega=2$ and $g=1$).     The initial coherent state parameter~$\zeta$ lies on a circle~$|\zeta|=1/2$.
    The evolution is plotted for various values of detuning~$\Delta$.
    Both $R(t)$ and $z(t)$ are periodic functions of time.
    }
    \label{fig:SU2TimeDependenceRadius}
\end{figure}

\begin{figure}[h]
    \begin{center}
        \includegraphics[width=0.47\textwidth]{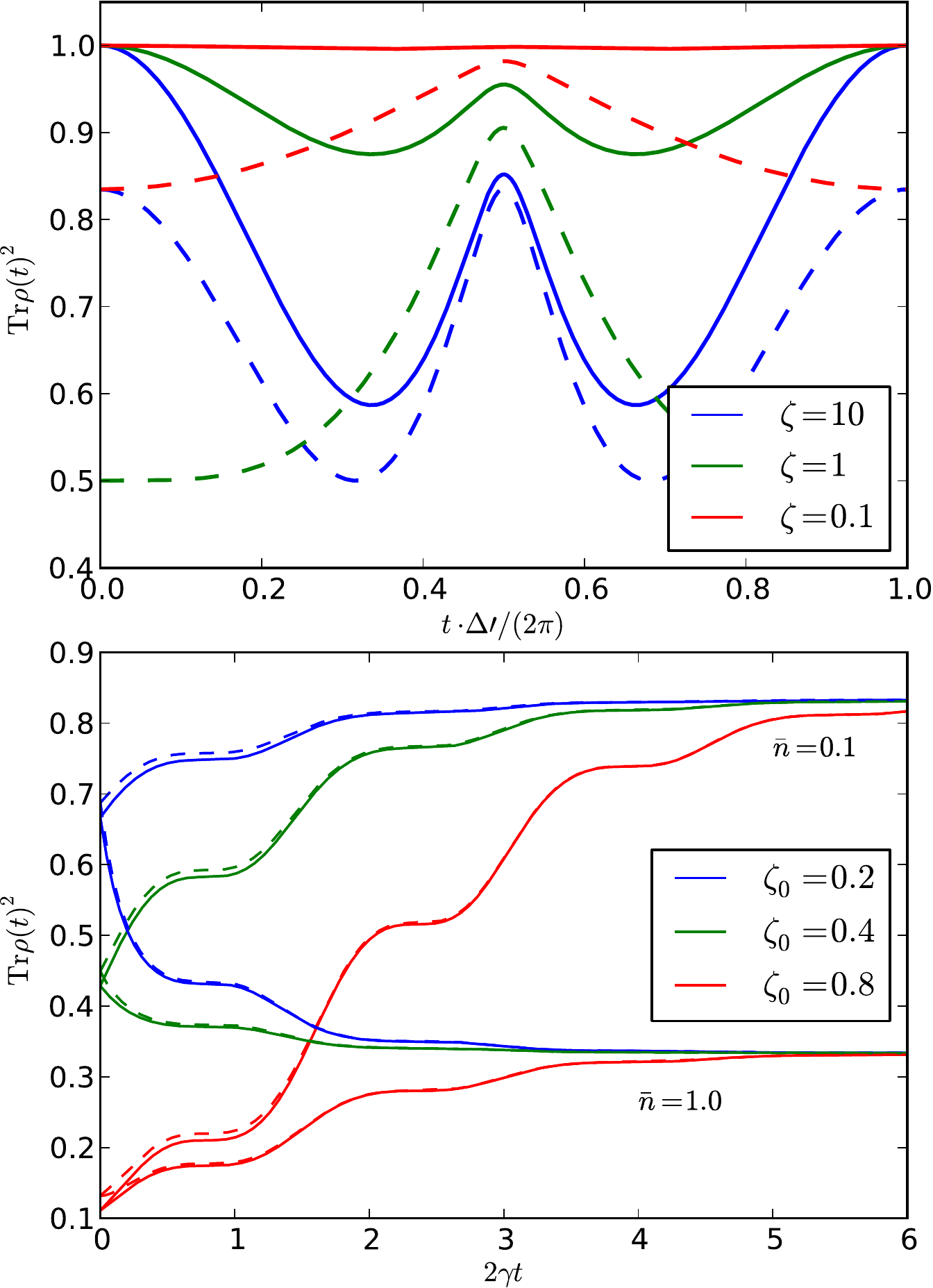}
    \end{center}
\caption{
The time evolution of the purity~$\Tr\rho^2(t)$ of the density matrix~$\rho$ for
a $su(2)$-type system (upper panel) defined by~Eq.~\eqref{eq:SU2Xt},
and a $su(1,1)$-type system (lower panel).  The initial state of the evolution is
a LCS with initial parameter~$\zeta$ (resp. $\zeta_0$) with different initial
purities. 
The $su(1,1)$ system is characterized by 
time-dependent decay rates
$\gamma_{1,2}(t) = \gamma (1+\cos(8 \gamma t))/2$. 
Full lines
describe the evolution when $c_m=0$  in
Eq.~\eqref{eq:SU11GeneralDensityMatrix} for $m>0$ (diagonal matrix).
Dashed lines
involve a small admixture of a coherent state~$\ket{1,\zeta_1}$.  
}
\label{fig:Purity}
\end{figure}

Thanks to the absence of dissipation (the bath has one mode only), $f_-(t)$~is  a
periodic function of time and therefore the coherent state parameter is periodic as well. 
As mentioned above, a circle in the plane~ of the initial coherent state
parameter~$\zeta(0)$ is mapped onto a circle in the plane of coherent state parameter~$\zeta(t)$ 
at time~$t$.
The ``pulsation'' of the parameters $R$ and $z$ of Eq.~\eqref{eq:CoherentCircle} is
plotted in Figure~\ref{fig:SU2TimeDependenceRadius}, while the time evolution of the purity of the density matrix is shown in
Fig.~\ref{fig:Purity}.

\section{Harmonic oscillator-type models: SU(1,1) LCS}
In analogy with the LCS for spin-boson type models, we can study models of a harmonic oscillator type,
which lead to a non-compact symmetry group~SU(1,1). 
A~generic model of a harmonic oscillator in contact with a reservoir is described by
the Lindblad  equation of the form~\eqref{LindEq} (c.f.~\cite{Puri}) with $H(t) = \omega(t)
a^\dagger a$, $A_+ = a^\dagger$, $A_- = a$, $\gamma_+ = \gamma_2(t) {\bar n}$ and $\gamma_- =
\gamma_1(t) ( {\bar n} +1 )$,
where~$\bar{n}$ is the equilibrium thermal occupancy of the oscillator and couplings
$\gamma_{1,2}(t)$~are system-specific. To proceed 
we denote $\hat{n}=a^{\dag}a$
and introduce the following superoperators:
%$K_{-}\rho = a\rho a^{\dag}$, $K_{+}\rho=a^{\dag}\rho a$, 
%$K_{0} = \frac{1}{2}(\hat{n}\rho+\rho \hat{n} +\rho)$, 
\begin{gather}
K_{-}\rho = a\rho a^{\dag}, \qquad K_{+}\rho=a^{\dag}\rho a\nonumber\\
K_{0}\rho = \frac{1}{2}(\hat{n}\rho+\rho \hat{n} +\rho), 
\quad R \rho = \hat{n} \rho - \rho \hat{n},
\end{gather}
which satisfy the~$su(1,1)$ commutation relations
\begin{gather}
[K_{-},K_{+}] = 2K_{0},\quad [K_{0},K_{\pm}] =\pm
K_{\pm}\nonumber\\ [R,K_\alpha]=0.
\end{gather}
The Casimir invariant is given by
$[K^{2}_{0}-1/2(K_{+}K_{-}+K_{-}K_{+})]\rho=\frac{1}{4}(-\rho+\hat{n}^{2}\rho+\rho
\hat{n}^{2} - 2 \hat{n} \rho \hat{n})$.  
The algebraic structure in this case is~$su(1,1)\times u(1)$ and we may again disantagle
the evolution operator using the method presented above. 
The evolution equation~$\dot\rho(t)=\mathcal{L}\rho(t)$ with
\begin{multline}
{\cal L} = [\gamma_1(t) ( {\bar n} + 1) -\gamma_2(t) {\bar n} ]
    -i\omega(t) R  \\
    + 2\gamma_1(t) ({\bar n} + 1 ) K_- 
    + 2 \gamma_2(t) {\bar n} K_+ \\
    - 2 [ \gamma_1(t) ({\bar n} + 1 ) +  \gamma_2(t) {\bar n} ] K_0
\end{multline}
is solved by 
\begin{multline}
\rho(t) = e^{f_+ K_+} e^{f_0 K_0} e^{f_- K_-}\\
e^{-i R \int \omega dt} e^{\int [\gamma_1(t) ( {\bar n} + 1) -\gamma_2(t) {\bar n} ]
dt } \rho(0),
\end{multline}
where $f_{\pm,0}$ satisfy Eq.~\eqref{eq:Ric} for $\sigma=1$.

The  space of density matrices can be decomposed into a direct sum of irreducible subspaces of the
underlying $su(1,1)$ algebra. 
Namely, the operators of the form $\ketbra{n+m}{n}$ (or~$\ketbra{n}{n+m}$) belong to an irreducible
subspace, according to  the action of the Casimir operator on them 
$\vec{K}^2 \ketbra{n+m}{n}  = k(k-1)\ketbra{n+m}{n}$ with $k=\frac{1}{2}\left( 1 +  m
\right)$ (similarly for~$\ketbra{n}{n+m}$).
Thus the space of density matrices is decomposed into 
a direct sum of subspaces, corresponding to discrete-series representations of the $su(1,1)$~algebra. 
We define~$\ket{m;n}$ as $\ketbra{m+n}{n}$ for~$m\ge0$ and $\ketbra{n}{n-m}$ for
$m<0$; then $K_0 \ket{m;n} = ( k + n ) \ket{m;n}$, with $n\ge 0$. The evolution operator acts
independently on each of these subspaces.

The coherent states are
introduced separately for each irreducible subspace~$m$ according to the recipe presented above in
this paper. Thus we parametrize each subspace~$m$ by coherent states~$\ket{m;\zeta_m}$
($\zeta_m$~being the coherent state parameter) built upon
the basis state~$\ket{m;0}$. 
A general density matrix  can be expressed in terms of the
over-complete basis of these coherent states as
\begin{equation}
    \rho(t) = \sum_{m\ge 0} \int d \mu_m(\zeta) \left[
    c_m(\zeta)  \ket{m;\zeta} +
    h.c.\right]
    \label{eq:SU11GeneralDensityMatrix}
\end{equation}
where~$c_m(\zeta)$~are appropriate time-dependent functions and $d\mu_m$~is the
group measure~(see Eq.~\eqref{eq:ResolutionOfUnity}). The coherent state parameters~$\zeta_m$ are all time-dependent; 
their evolution is given by Eq.~\eqref{eq:CoherentStateEvolution}.
In particular, we can consider a density matrix being a purely coherent state within each
$m$-sector.
There are, however,  constraints on the admissible set of
parameters~$c_m,\zeta_m$, in order for the linear combination to indeed represent
a physical density matrix.  The simplest one reads~$1 = \Tr\rho =
(1-|\zeta_0|^2)^{-1/2}[ c_0/(1-\zeta_0) + c.c.]$. Other constraints,  such as $0
\le \Tr\rho^2 \le 1$, are more involved; as a result not every purely coherent
state is physically allowed. Interestingly, the decomposition of an arbitrary
physical density matrix  may contain even these unphysical states. Because of
this feature no simple parametrization of a coherent state in terms of its
purity exists, contrary to the $su(2)$~case.

\begin{figure}
    \begin{center}
        \includegraphics[width=0.47\textwidth]{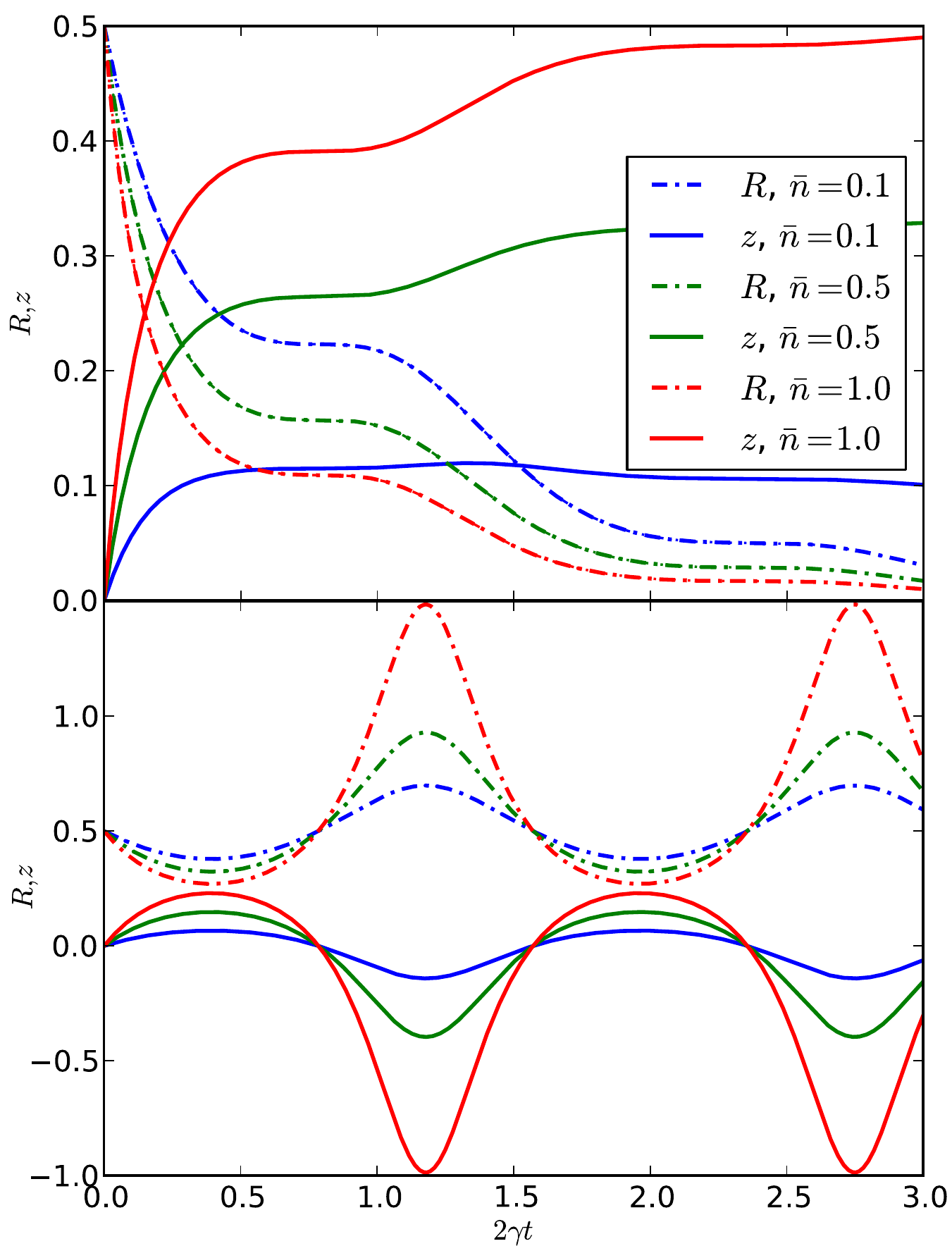}
    \end{center}
    \caption{
    The time dependence of the parameters $R$ and $z$, introduced in
    Eq.~\eqref{eq:CoherentCircle}, for the $su(1,1)$-type model defined by 
    Eq.~\eqref{eq:SU11GammaTimeDep}.
    The upper panel shows the evolution for the parameter~$a=1$, while the lower
    one for~$a=0$, for various values of the thermal occupancy~$\bar{n}$.
    }
    \label{fig:SU11GammaPeriodicGammaRadiusEvol}
\end{figure}

In order to demonstrate the LCS  technique in the $su(1,1)$~case on a particular
example, we focus on a harmonic oscillator interacting  with a bath in such a
way that the decay rates $\gamma_{1,2}(t)$ are both equal to a periodic
function
\begin{equation}
    \gamma_1(t) = \gamma_2(t) = \gamma(t) = \gamma ( a + \cos \Gamma t ),
    \label{eq:SU11GammaTimeDep}
\end{equation}
where~$a$ is a free
parameter.  The time evolution of the parameters $R$ and $z$ of
Eq.~\eqref{eq:CoherentCircle} can be seen in
Fig.~\ref{fig:SU11GammaPeriodicGammaRadiusEvol}, while
the evolution of the purity~$\Tr\rho(t)^2$ is captured in Fig.~\ref{fig:Purity}.

\section{Conclusion} 
In this paper we introduced the notion of Liouville coherent
states, an analogue of generalized coherent states for the density matrices.
This concept identifies a class of density matrices whose time evolution is
robust with respect to any time-dependent driving: if the initial density matrix
belongs to this class it will remain so for any $t>0$. We demonstrated this on
several physical examples, involving compact and non-compact dynamical symmetry
groups. Many geometric and
topological properties of these states deserve further study. As these states
form an overcomplete basis, they can be used as a platform for investigating a more
 complicated Liouville dynamics analogously  to the way in which the usual
 coherent states
are used in path-integral formulation of the quantum dynamics.

\end{document}